\documentclass[twocolumn,showpacs]{revtex4}
\usepackage{graphicx}

\begin{document}

\title{The Arrow of Time and Correlations in Quantum Physics}
\author{Vlatko Vedral}
\affiliation{Clarendon Laboratory, University of Oxford, Parks Road, Oxford OX1 3PU, United Kingdom and\\Centre for Quantum Technologies, National University of Singapore, 3 Science Drive 2, Singapore 117543 and\\
Department of Physics, National University of Singapore, 2 Science Drive 3, Singapore 117542\\Center for Quantum Information, Institute for Interdisciplinary
Information Sciences, Tsinghua University, Beijing, 100084, China}

\begin{abstract}
We discuss the arrow of time in terms of the increase of correlations between the system and its environment. Here we show that the existence of the arrow of time, based on deleting correlations, requires a strict absence of initial correlations between the system and the environment. We discuss our work in relation to other approaches addressing the same problem and emphasise similarities and differences.   
\end{abstract}

\maketitle

{\bf Introduction.} Irreversible processes are ubiquitous in nature, but they present a deep mystery to a physicist. The most fundamental dynamical laws, those of quantum physics, are fully reversible in time (this is true of other types of dynamics, such as Newtonian and electrodynamics, but here we will confine our analysis to quantum physics). The only way in which irreversibility can arise out of reversible dynamical laws is through imposing additional restrictions. Broadly speaking, these come in two different guises. Either we restrict the set of operation available to us, or we restrict the set of allowed states that the system can occupy (or both, of course). For instance, in thermodynamics we restrict ourselves to adiabatic transformations, in which case the only allowed processes are those that increase entropy. Without restrictions, the entropy of any system can also decrease (unless the system is closed, in which case the entropy remains constant). To take a more modern example, we restrict ourselves to local operations and classical communication, and the result is that entanglement of a multipartite system typically decreases \cite{Vedral-RMP}. If, instead, operations are allowed to be global, entanglement can also increase and so we lose unidirectionality. Restricting operations further by allowing only local ones without classical communication, we notice that total correlations, as measured by the mutual information, now also monotonically decrease. This is a very important property which justifies why we use mutual information to quantify correlations \cite{Vedral-PRL}. It is the changes in mutual information that will feature prominently in what follows. 

The arrow of time is linked with irreversibility \cite{Gold}, which itself is manifested by the increase in entropy of an isolated system. Of course, in quantum physics an isolated system evolves unitarily and its entropy can therefore not change in time. It is sometimes claimed that measurements are responsible for irreversibility and temporal asymmetry. It is true that the entropy of the state suffering a complete projective measurement increases, but we can always phrase measurements in a temporally symmetric fashion \cite{Aharonov}. Here the initial and final states of the system are fixed, and we deal with the probability for a measurement outcome in between. The resulting formula, which can be taken as the basis for quantum physics, is completely symmetric under reversing the initial and the final boundary condition. This is basically because $|\langle m |P|n\rangle|^2 = |\langle n |P|m\rangle|^2$. Therefore, it seems that not even quantum measurements can be responsible for the arrow of time. We will shortly see that there is another way of reaching the same conclusion. 

It appears that the only way to increase entropy is, as we said, to impose restrictions. Imagine, however, that instead of restricting operations, we restrict the overall states to local states and neglect any correlations (a procedure akin to ``coarse graining", though, strictly speaking, the term coarse graining implies looking at averages of quantities). The irreversibility then arises due to this neglect of correlations as can be seen from the following argument \cite{Peres,Everett}. Suppose we start with the total state of the universe of the form: $\rho_S\otimes\rho_R$. Assume now a general global unitary evolution. The final state is: $\rho'_{SR}=U_{SR}\rho_S\otimes\rho_RU^{\dagger}_{SR}$. The change of entropy in the system plus rest is
\begin{eqnarray}
\Delta S_S + \Delta S_R & = & (S( \rho'_{S}) - S( \rho_{S})) + (S( \rho'_{R}) - S( \rho_{R}))\nonumber \\
& = & S( \rho'_{S}) + S( \rho'_{R}) - S(\rho_{SR}) \nonumber \\
& = & S( \rho'_{S}) + S( \rho'_{R}) - S(\rho'_{SR}) \nonumber \\
& = & I(S':R')\geq 0 \nonumber
\end{eqnarray}
where $I(S:R)= S( \rho_{S}) + S( \rho_{R}) - S(\rho_{SR})$ is the von Neumann mutual information between the system and the rest. The last inequality follows from the fact that the mutual information is non-negative which itself follows from the subadditivity of entropy. The increase in local entropies of the system and the rest is exactly equal to the mutual information in the final state between the system and the rest (provided they started out as completely uncorrelated). Therefore, erasing correlations leads to entropy increase.

Note that we do not lose anything by looking at global unitaries only and excluding measurements. Measurements themselves obviously change the entropy of the system (see \cite{Partovi-I,Vedral}), but they could always be presented as a unitary transformation on a larger system. Given that $R$ includes the rest of the universe here, this possibility is automatically covered. This is another way of understanding why quantum measurements cannot explain irreversibility (unless the measurement itself is for some reason fundamentally irreversible).   

Ignoring all initial correlation seems like a brutal restriction (it is sometimes called the ``assumption of molecular chaos", a phrase coined by Boltzmann when he was wrestling with explaining irreversibility). Is there any physical basis for this? Boltzmann reasoned that any two colliding molecules are highly improbable to collide again. A modern way of encapsulating Boltzmann's logic is by using thermalizing quantum machines \cite{Scarani}.  Here the system is a qubit (though it could be more general in principle), and it approaches equilibrium after a number of successive two-qubit interactions with qubits of the reservoir. Because each new interaction is with a different reservoir qubit, the state (of the two qubits about to interact) is always completely uncorrelated. The irreversibility here comes simply from the complexity of reversing all the collisions (which could in principle be done, but it requires us to keep tract of the exact order in which they take place).  

But at the level of the universe we can perhaps argue more generally. Different parts of the universe which might initially be interacting, could then undergo a rapid expansion leading to them becoming causally disconnected. This is the view of inflation, which is currently predominant in cosmology. In that case, correlations that might initially have been established are naturally erased during the inflationary period. Other processes, such as black hole formation can also serve as causal horizons and could therefore be seen as naturally correlation deleting (rather than being carelessly ignored).  

Ignoring correlations between the system and the rest of the universe can be based on less esoteric phenomena, namely the fact that the rest of the universe is of very low entropy (this happens to be true for our universe!). Then the system and the rest cannot be very correlated to each other at any instant in time (since the rest of the universe is in a near pure state, i.e. at very low temperatures). Namely, the mutual information is in this case very close to the entropy of the rest itself. But is this existing low level of correlations still a problem for the directionality of entropy? 

It is clear that if the initial state is not uncorrelated, the resulting change of entropy can indeed be negative. A simple example involves an initial maximally entangled state \cite{Partovi-II,Rudolph}. 

{\bf Example}. Suppose that the initially maximally entangled state evolves into a product state. Then the initial entropy is maximal, while the final is zero. Therefore we have a decrease in local entropies. 

It is also clear that we do not need an entangled starting state in order to achieve a decrease in the sum of local entropies\cite{Rudolph}. Suppose that we have just a classically correlated initial state. Then, consider a unitary transformation that decorelates the system from the rest, i.e.
\begin{eqnarray}
& & \frac{1}{2} \{|0_S0_R\rangle \langle 0_S0_R| + |1_S1_R\rangle \langle 1_S1_R|\} \rightarrow \\ 
& & \frac{1}{2} \{|0\rangle \langle 0|_S + |1\rangle \langle 1|_S\} \otimes |0\rangle\langle 0|_R 
\end{eqnarray} 
Here the local entropy decreases too. But the situation is, in fact, more problematic than this. We now prove the main result of this paper: not only can initial classical correlations lead to an entropy decrease, but we show that any small departure from the product state (irrespectively of whether this generates entanglement or not) can always lead to entropy reduction. 

{\bf Main result}. In the vicinity of any product state (meaning arbitrarily close to the product state as measured by, say, the Bures metric), there is a state whose mutual information can be reduced by a global unitary transformation.

{\bf Proof}. Let us start with a state that is close to a product state
\begin{equation}
\rho_{SR} = (1-\epsilon) (|0_S0_R\rangle\langle 0_S0_R|) + \epsilon |\Psi^+_{SR}\rangle\langle \Psi^+_{SR}|  
\end{equation}
where $|\Psi^+_{SR}\rangle = |01\rangle + |10\rangle$. This state clearly has a non-zero mutual information. If we now exectue a unitary transformation such that 
$|\Psi^+_{SR}\rangle \rightarrow |01\rangle$, while the state $|00\rangle$ remains the same, then the final state will be a product state between $S$ and $R$ and therefore
the mutual information will be decreased.

If the starting state is more generic, such that we are close to a product of mixed states, $\rho_S\otimes \rho_R$, we can always represent it as a product of pure states at the higher level (by purifying the system and the rest). Then the argument is the same as already presented.   

{\bf Corollary.} The sum of local entropies can always be made to decrease if the state is not a product state. 

This, of course, is not a very satisfactory conclusion. It means that unless we have a strict initial product state, there is always a possibility of decreasing the sum of local entropies. If the starting state is uncorrelated, then any transformation can only increase correlations and this is then what gives us the arrow of time. In practice, therefore, this way of arguing for irreversibility is not useful since we effectively never have a product state between the system and the rest of the universe unless the rest is strictly speaking at zero temperature. Note again that entanglement has nothing to do with this, not even entanglement within the rest of the universe (one could argue that if the rest is very entangled then it cannot be strongly correlated to the system). If the rest of the universe has a certain finite number of qubits (the current estimate for total number of qubits in the universe is about $10^{122}$) then we can always choose to be close enough to a maximally mixed state that entanglement is guaranteed not to exist.  

{\bf Schr\"odinger's attempt at rescue}. Let us now consider another possible way of reconciling this entropy decrease \cite{Schrodinger}. Intuitively, it rests on the fact that there is no absolute arrow of time in the whole universe (since it is a closed system). What matters, on the other hand, is the relative arrow direction of the system compared with that of the rest of the universe. Namely, we would like both to be pointing in the same direction.
The reasoning behind this is that system's entropy could go down, signifying the reversal of the arrow of time, but if the rest of the universe also goes down in entropy (i.e. reverses its own arrow of time), then we have no inconsistency (it is perfectly feasible that this happens in our universe, but we would not notice anything wrong because every time we become younger, so does the rest of the universe!). So it is possible for entropy to decrease, but there is no problem (according to this logic) if this is followed by the decrease in the rest of the universe, or, in other words, we need $\Delta S_S \times \Delta S_R \geq 0$ to hold. This was exactly the point Schr\"odinger himself made in one of his more entertaining accounts of the arrow of time \cite{Schrodinger} (for a similar recent attempt see \cite{Maccone}). Unfortunately, within our setting, this version of the second law can also be violated. Under the assumptions so far, the entropy of the system can evolve in the opposite direction of the rest of the universe. The key question to ask then is if this possibility can be naturally eliminated with some extra assumptions hitherto uninvoked? 

{\bf Discussion}. We have been looking at restricting physical states of the universe. Now we turn to imposing additional restrictions, but on the unitary evolution itself. Do we have any sound physical basis for doing so? One such restriction is to allow only weak couplings between the system and the rest. This means that the strength of the coupling Hamiltonian is small enough that we can write the unitary evolution as $U\approx I - i H t$ (sometimes known as the Born approximation). Here it is clear that we will not be able to decrease entropy for any state in the vicinity of a product state. Another way of recovering the individual entropy increase is to argue that when the system and the rest have low correlations, most global unitaries will tend to increase correlations (so the arrow of time is statistically very likely to be in the direction of individual entropy increase).  

A similar restriction on both product states as well as transformations was assumed by Partovi\cite{Partovi} and Reents \cite{Reents} (see also \cite{Jevtic-1,Jevtic-2}) in order to argue for the quantum derivation of the fact that heat always flows from the hot to the cold system. Reents assumed that the starting state of two subsystems is a product of the Gibbs states for the system and the rest and that the unitary evolution is governed by a special Hamiltonian (namely an interaction Hamiltonian that commutes with the Hamiltonians of the individual subsystems, \cite{Jevtic-1,Jevtic-2}). Our considerations are more general (in that we do not need Gibbs states as starting or final states and the evolution can be any global unitary transformation). The link with Reents can still be made as follows. Suppose that we define the temperature of the system as $T_{S} = \Delta U_S/\Delta S_S$ and likewise for the rest (note that given that our initial states need not be Gibbs, this would be an ``effective" temperature). Then, the fact that $\Delta S_S + \Delta S_R \geq 0$ implies that
\begin{equation}
\frac{\Delta U_S}{T_S} + \frac{\Delta U_R}{T_R} \geq 0
\end{equation}  
This itself means that if $T_R < T_S$ then $\Delta U_R > \Delta U_S$, i.e. the heat flows from the hotter to the colder of the two subsystems (the latter assumed to be the rest of the universe in line with the above discussion). This is, of course, another form of the second law, namely the fact that heat always flows from the hot (i.e. system) to the cold object (i.e. the rest). 

It is worth noting that in order to communicate the main message (i.e. the need for the strict initial product state) we have (deliberately, so far) ignored one more possibly important issue. Namely the rest of the universe could be divided into many subsystems (as with thermalizing quantum machines) and they could all have different arrows of time (not the case with thermalizing machines since the system qubit and the reservoir qubit coupling is always the same). The possibility of different arrows of time in different parts of the universe has already been pointed out a number of times (see e.g. \cite{Rudolph,Page}) and it only increases the strength of our conclusion: in order to have the same arrow of time as defined here all of subsystems need to be completely uncorrelated.  

Finally, we note that the Partovi and Reents assumptions are related to another approach to irreversibility, proposed by Crooks \cite{Crooks} and based on the Jarzynski equality\cite{Jarzynski,Crooks}. Here the scenario is that we start from a Gibbs state of the system and the rest, $\rho_R\otimes \rho_S$, just like in Partovi and Reents. But unlike them, we now proceed by first making projective measurements in the respective eigenbasis of $S$ and $R$, $P^n_{SR}$, where $n$ labels the $n$-th eigenstate. A generic unitary transformation is then applied (again, Reents has a restriction here too), followed by another measurement $Q^m_{SR}$, but this time in the basis of the final Hamiltonian, $H_f$ (this basis could be an entangled one between the system and the rest). The probability to obtain the initial outcome $n$ followed by the final outcome $m$ is given by
\[ 
p_f(n,m) = tr (Q^m UP^nU^{\dagger}) p_n \; ,
\]
where $f$ stands for ``forward" (and I have dropped the subscripts $SR$) and $p_n$ is the initial Gibbs distribution (jointly between the system and the rest). The backward distribution $p_b(n,m)$ is defined by reversing this protocol in that we state from the thermal state of $H_f$, then make measurements $Q^m$. This is then followed by the evolution $U^{\dagger}$ and finally by the measurement $P^n$. Therefore 
\[ 
p_b(n,m) = tr  (P^n U^{\dagger}Q^mU) q_m \; ,
\]
where $q_m$ is the Gibbs distribution on $H_f$. Dividing the two probabilities, we obtain
\[ 
\frac{p_f(n,m)}{p_b(n,m)} = e^{\beta (W_{nm} - \delta F)} \; .
\]
where $W_{nm}$ is the difference between the initial energy eigenvalue and the final one and $\delta F$ is the free energy difference between the Gibbs states on $H_i$ and $H_f$. Taking the natural logarithm of both sides and averaging over with $p_f(n,m)$ we obtain
\[ 
S(p_f(n,m)||p_b(n,m)) = \sum_{n,m} p_f(n,m) S_{nm} \; ,
\]
where $S_{nm} = W_{nm} - \delta F)$ is and the right hand side is therefore an average entropy increase in the forward direction that is simply equal to the relative entropy between the forward and backward probability distribution. However, this is not the same as the mutual information of the final state as appears in the previous approach. A simple way of seeing that is to assume that the final state is also a product state. Then, in general, $\sum_{n,m} p_f(n,m) S_{nm} > 0$, while $I = 0$ (since no correlations are generated).  The origin of temporal asymmetry is therefore different in the Crooks approach. It comes from the fact that the reverse protocol starts from the thermal state of the final Hamiltonian in the forward direction (instead of starting from the state $U\rho_S\otimes \rho_RU^{\dagger}$). Note that the measurements on their own would not suffice since, as we remarked earlier, $p(m|n) = tr (Q_m UP_nU^{\dagger}) = tr  (P_n U^{\dagger}Q_mU) = p(m|n)$. In that sense, we are talking about the heat increase when the state $U\rho_S\otimes \rho_RU^{\dagger}$ is damped into the reservoir with state $e^{-\beta H_f}/Z_f$ (which is just given by the relative entropy between the two states \cite{Leff}). It is the thermalisation that leads to irreversibility, rather than deletion of correlation (which may, or may not, take place during the process of thermalisation). 

In summary, here we treat the arrow of time as a manifestation of the increase of correlations, as quantified by the mutual information, between the system and the rest of the universe. It is not, by any means, clear that this is the appropriate way of explaining the arrow of time (or that, indeed, that there is a way of explaining the arrow of time). What is important, of course, is that this suggestion can be falsified. This is based on our main result, namely that the presence of initial correlations could always in principle lead to the entropy decrease within this scenario. We can perfectly imagine a future time where the universe is much more correlated then now (since, as far as our current observations suggest, objects like stars will be cooling down to ultimately reach the temperature of the cosmic background). At some stage, therefore, even though the coupling may still be weak, the correlations between systems and environments may be big enough that the entropy might be decreased through their unitary interaction. This would, presumably, imply that violations of the Second law (if its explanation is based on the understanding exposed here) would become observable and that we would be able to gain work without having to waste any heat. The resource used will then simply be the initial correlations between the system and the rest.

\textit{Acknowledgments}: The author acknowledges funding from the John Templeton Foundation,
the National Research Foundation (Singapore), the Ministry of Education (Singapore), the Engineering
and Physical Sciences Research Council (UK), the Leverhulme Trust, the Oxford Martin
School, and Wolfson College, University of Oxford. This research is also supported by the National
Research Foundation, Prime Ministers Office, Singapore under its Competitive Research
Programme (CRP Award No. NRF- CRP14-2014-02) and administered by Centre for Quantum
Technologies, National University of Singapore.

\end{document}